\documentclass[twocolumn,prb,showpacs]{revtex4}
\usepackage{graphics}
\usepackage{graphicx}
\usepackage{subfigure}
\usepackage{longtable}
\usepackage{multirow}
\usepackage{bm}

\graphicspath{{.}}

\begin{document}
\title{First-principles calculations of graphene nanoribbons in
  gaseous environments: Structural and electronic properties}
\author{M. Vanin} \affiliation{Center for Atomic-scale Materials
  Design, Department of Physics \\ Technical University of Denmark, DK
  - 2800 Kgs. Lyngby, Denmark} 
\author{J. Gath} \affiliation{Center
  for Atomic-scale Materials Design, Department of Physics \\
  Technical University of Denmark, DK - 2800 Kgs. Lyngby, Denmark}
\author{K. S. Thygesen} \affiliation{Center for Atomic-scale Materials
  Design, Department of Physics \\ Technical University of Denmark, DK
  - 2800 Kgs. Lyngby, Denmark} 
\author{K. W. Jacobsen}
\affiliation{Center for Atomic-scale Materials Design, Department of
  Physics \\ Technical University of Denmark, DK - 2800 Kgs. Lyngby,
  Denmark}

\date{\today}

\begin{abstract}
  The stability of graphene nanoribbons in the presence of typical
  atmospheric molecules is systematically investigated by means of
  density functional theory. We calculate the edge formation free
  energy of five different edge configurations passivated by H, H$_2$,
  O, O$_2$, N$_2$, CO, CO$_2$, and H$_2$O, respectively. In addition
  to the well known hydrogen passivated armchair and zig-zag edges, we
  find the edges saturated by oxygen atoms to be particularly stable
  under atmospheric conditions. Saturation of the zigzag edge by
  oxygen leads to the formation of metallic states strictly localized
  on the oxygen atoms.  Finally, the vibrational spectrum of the
  hydrogen and oxygen passivated ribbons are calculated and compared.
\end{abstract}

\pacs{73.20.At, 73.22.Pr, 71.15.Mb}
\maketitle

\section{Introduction}
Graphene, a single atomic carbon layer, has attracted extensive
research interests in the last few
years\cite{Novoselov:2004,Novoselov:2005,Geim:2007,Castro:2009}.
Recent advances in experimental techniques have lead to the
fabrication of several graphene-based structures for both fundamental
physics investigations and promising applications. Graphene
nanoribbons (GNRs), in particular, have received considerable
attention due to their possible role in future carbon-based
nanoelectronics\cite{Wang:2008,Li:2008}. GNRs are graphene strips of
finite width, in which lateral confinement opens an electronic gap as
opposed to the vanishing gap in infinite graphene sheets.

Theoretical calculations and experiments have shown that the
electronic and transport properties of GNRs are strongly influenced by
the actual atomic configuration of the edges \cite{Ritter:2009} which,
contrary to bulk graphene, are very reactive. A detailed knowledge of
the stability and electronic properties of the different types of
edges is therefore required in order to understand and ultimately
design GNRs with specific properties.

Most theoretical calculations on GNRs have considered the the armchair
and zigzag edges with or without hydrogen passivation.  However,
recent calculations have shown that other edge configurations can also
show high stability\cite{Koskinen:2008}. Since molecular hydrogen is
one of the most common contaminants even in ultra high vacuum
experiments, it is certainly relevant to consider its influence at the
edges\cite{Wassmann:2008}. On the other hand, especially at room
temperature and under ambient conditions, the effect of other gas
molecules on the edge stability should also be considered.  This is
also relevant from the perspective of chemical
functionalization\cite{Kan:2008,Cervantes:2008}.

\begin{figure}
    \includegraphics[width=0.5\textwidth]{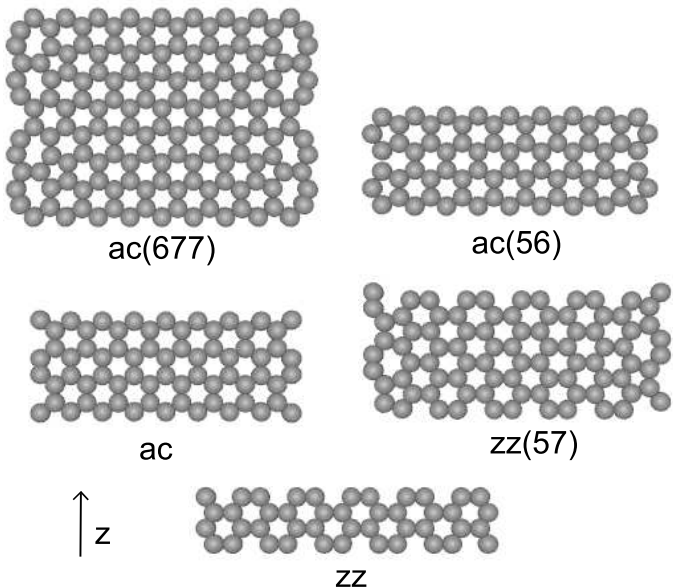}
    \caption{The standard and reconstructed edge geometries (and
      corresponding abbreviations) considered in this study. The
      figures show two unit cells of the nanoribbons with periodicity
      along the $z$-direction.}
  \label{fig:1}
\end{figure}

In this paper, we study the stability of five different edge
geometries in the presence of H$_2$, O$_2$, N$_2$, CO, CO$_2$ and
H$_2$O using first principles calculations. The different edge
geometries for the pristine ribbons with the corresponding
abbreviations are shown in Fig. \ref{fig:1}. We find that the edges
passivated by atomic oxygen are the most stable under atmospheric
conditions closely followed by the zig-zag and armchair edges with
hydrogen adsorbed. The oxygen passivated zig-zag GNR is found to be
metallic with the states at the Fermi level strictly localized on the
oxygen $p$ orbitals. The vibrational spectrum of the hydrogen and
oxygen passivated GNRs are calculated and discussed.  Throughout the
paper we focus on free-standing graphene and thus neglect any effect
related to the substrate. Whether this is a reasonable assumption
clearly depends on the strength of the substrate-graphene
interaction. This is likely to make the edges less reactive towards
gas species, since the edges would be already stabilized by the
chemical bond to the substrate. For metallic substrates the nature of
the substrate-graphene interaction is still unclear, but DFT
calculations indicate that it can vary considerably depending on the
metal\cite{Giovannetti:2008,Vanin:2010}.

\section{Computational details}
Density functional theory (DFT) calculations are performed with the
GPAW code\cite{Mortensen:2005,Jussi:2010} which is a real space
implementation of the the projector augmented wave
method\cite{Blochl:1994}. The widths of the nanoribbons are in the
range $20-23 \, \mathrm{\AA}$, corresponding to 10-12 carbon dimers,
which is needed in order to get fully converged results with respect
to the width. Note that this size is also relevant to modern
experiments, where sub-10nm wide GNRs are now
achieved\cite{Wang:2008,Li:2008}. The GNRs are separated by 10 \AA\ in
all directions. All calculations are spin-polarized and use the
RPBE\cite{RPBE:1999} exchange-correlation functional. A grid-spacing
of $0.18 \, \mathrm{\AA}$ is used, and a (1$\times$1$\times$10)
Monkhorst-Pack $k$-point grid is employed to sample the Brillouin zone
along the periodic direction of the nanoribbons. We relax all the
structures until the maximum force is lower than $0.05$ eV/\AA.

\section{Stability}
\subsection{Adsorption energies}
\begin{figure}
    \includegraphics[width=0.49\textwidth]{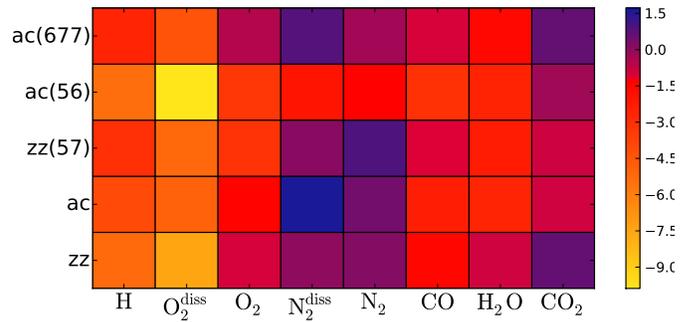}
    \caption{(Color online) Adsorption energy, $E_{\mathrm{ads}}$, for
      the different molecular species on the different edge
      configurations in eV. In the case of nitrogen and oxygen we have
      considered both the molecular and dissociated forms on the
      edge. In all cases the adsorption energy is given per
      \emph{molecule}.}
  \label{fig:2}
\end{figure}

For each of the five edge configurations we calculate the adsorption
energy of the different molecules according to
\begin{equation}
  \label{eq:2}
  E_{\mathrm{ads}} = E^{\mathrm{ribb+mol}} - N_{\mathrm{mol}} E^{\mathrm{mol}} - E^{\mathrm{ribb}} 
\end{equation}
where $E^{\mathrm{ribb+mol}}$ is the total energy of the ribbon with
the adsorbed gas molecule, $N_{\mathrm{mol}}$ is the number of
adsorbed molecules, $E^{\mathrm{mol}}$ is the total energy of the
molecule in the gas phase and $E^{\mathrm{ribb}}$ is the total energy
of the pristine ribbon. In order to avoid the erroneous DFT
description of the O$_2$ gas-phase triplet groundstate, we use
$\mathrm{H_2O}$ and $\mathrm{H_2}$ as suggested in
Ref. \onlinecite{Norskov:2004}. The adsorption energies are summarized
in Fig. \ref{fig:2} and shown in \ref{tab:table1}.

\begin{table}
  \begin{tabular}{c|c|c|c|c|c|c|c|c}
    \hline
    \hline
    &$\mathrm{H}$&$\mathrm{O_2^{diss}}$&$\mathrm{O_2}$&$\mathrm{N_2^{diss}}$&$\mathrm{N_2}$&$\mathrm{CO}$&$\mathrm{H_2O}$&$\mathrm{CO_2}$ \\
    \cline{1-9}
    ac(677)& -2.59& -4.33& -0.45& 0.86& -0.22& -0.97& -1.46& 0.66 \\
    \hline
    ac(56)& -5.36& -9.88& -3.22& -1.95& -1.25& -3.03& -2.49& -0.20 \\
    \hline
    zz(57)& -2.99& -5.13& -3.09& 0.14& 0.96& -1.05& -2.24& -0.82 \\
    \hline
    ac& -3.99& -4.90& -1.29& 1.73& 0.44& -2.31& -2.56& -0.86 \\
    \hline
    zz& -5.19& -7.44& -0.96& 0.05& 0.20& -1.42& -0.85& 0.65 \\
    \hline
    \hline
  \end{tabular}
  \caption{Numerical values of the adsorption energies, $E_{\mathrm{ads}}$, corresponding to Fig. \ref{fig:2}. All values are in eV.}
  \label{tab:table1}
\end{table}

It can be seen that saturation by oxygen is the most favorable for all
edge configurations.  We note that this is in agreement to Ref.
\onlinecite{Lee:2009} in which the authors conclude that zig-zag
nanoribbons are more likely to be oxidized than hydrogenated. 

While the high reactivity of the zig-zag edge is well
estabilished\cite{Jiang:2007}, we also find the ac(56) to be very
reactive; in fact for several species this is the most reactive edge
configuration. As for the zig-zag edge, the origin of the high
reactivity of the ac(56) edge is a (spin polarized) peak in the density
of states close to the Fermi energy. Consistent with
their high reactivities, the pristine zz and ac(56) edges also have the
highest edge formation energies (see Table \ref{tab:table2}), i.e.
they are the least stable edges in the absence of molecular gas
species.

Finally, we note that all the reported results are for adsorption at
the edge carbon atoms since these are by far the most reactive
sites. For example we find that $\mathrm{H}$ adsorption on a carbon
atom located next to the edge is already unfavorable compared to
$\mathrm{H_2}$ in the gas phase.

\subsection{Edge formation energies}
The adsorption energies discussed in the previous section are relevant
once a given edge configuration has been formed.  More generally, to
thermodynamic stability of an edge in the presence of gas molecules
should also take the energetic cost of forming the edge into account.
To calculate the edge formation free-energy we first calculate the
formation energy of the pristine edges,
\begin{equation}
  \label{eq:1}
  E_{\mathrm{f}}=\frac{1}{2L} \left( E^{\mathrm{ribb}}-N_{\text{C}}E^{\mathrm{bulk}} \right)
\end{equation}
where $E^{\mathrm{ribb}}$ is the total energy of a ribbon with
$N_{\text{C}}$ carbon atoms in the supercell and $E^{\mathrm{bulk}}$
is the total energy per atom in bulk graphene. Our calculated energies
shown in Table \ref{tab:table2} are in very good agreement with
Ref. \onlinecite{Koskinen:2008}. We remark, that for pristine GNRs the
zz(57) reconstruction has the lowest edge energy.

The adsorption energy $E_{\mathrm{ads}}$ is used to obtain a
free-energy by taking into account the chemical potential,
$\mu_{\mathrm{gas}}$, of the gas at a given temperature and pressure
\begin{equation}
  \label{eq:3}
  G_{\mathrm{ads}} = \frac{1}{L} E_{\mathrm{ads}} - \rho_{\mathrm{gas}} \mu_{\mathrm{gas}} 
\end{equation}
where $\rho_{\mathrm{gas}}=N_{\mathrm{mol}}/L$ is the density of
adsorbed species. The chemical potential at temperature $T$ and
partial pressure $P$ is given by
\begin{equation}
  \label{eq:4}
  \mu = H^0(T) - H^0(0) - TS^0(T) + k_BT \mathrm{ln} \left( \frac{P}{P^0} \right)
\end{equation}
where $H^0$ and $S^0$ are enthalpies and entropies of the gas phase
molecules at $P^0=1$ bar.

Finally, the edge formation free-energy in the presence of adsorbates
is obtained as
\begin{equation}
  \label{eq:5}
  G_{\mathrm{f,ads}} = E_{\mathrm{f}} + G_{\mathrm{ads}}
\end{equation}
The vibrational contributions to
$G_{\mathrm{f,ads}}$ are not included since they are expected to be
insignificant\cite{Wassmann:2009}.

\begin{figure}
    \includegraphics[width=0.50\textwidth]{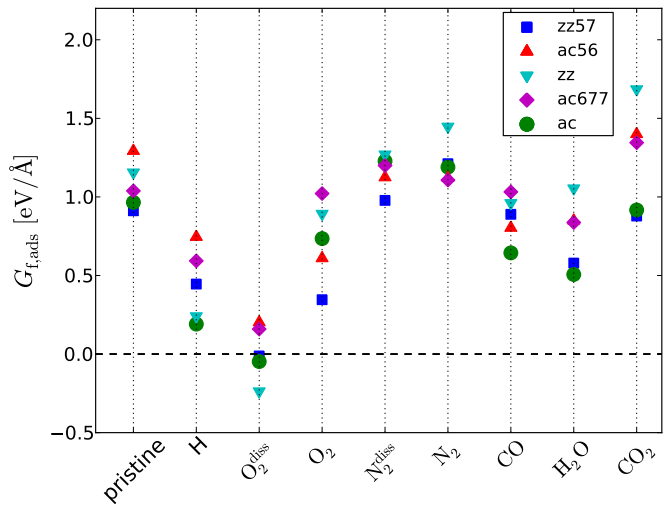}
    \caption{(Color online) Edge formation free energy of graphene
      nanoribbons with different edge structures (different symbols)
      and different adsorbed molecules at the edge. The free energy is
      evaluated at $300K$. The pressure for each structure with a
      given adsorbate is taken to be the partial pressure of the
      gas-phase adsorbate at atmospheric conditions.}
  \label{fig:3}
\end{figure}

Fig. \ref{fig:3} summarizes the calculated edge formation
free-energies at $300 K$ and atmospheric partial pressures for the
different gas species. The qualitative picture does not change if the
temperature and pressure ranges are varied within experimentally
accessible ranges. The results are also reported in Table
\ref{tab:table2} together with the edge formation energies calculated
at $0K$. It is immediately apparent that edges saturated by atomic
oxygen are the most stable. In particular, the lowest energy
configurations are the standard zigzag and armchair, followed by the
reconstructed zz(57); the other reconstructed edges are generally less
stable. This agrees with Ref. \onlinecite{Wassmann:2008} where
hydrogen passivation was considered.  The armchair and zigzag edges
saturated by hydrogen are also quite stable thus making $\mathrm{H}$
and $\mathrm{O}$ saturations the most stable ones. For this reason we
will focus in particular on the oxygen saturated GNR, i.e. the most
stable one, in the rest of the paper. Note that the negative formation
free-energies for some of the oxygen saturated edges mean that a
graphene sheet will spontaneously break in order to generate edges and
lower its energy. However one needs to be careful in analyzing these
results, since significant energy barriers may be involved in the
process and they are not included in our study. Furthermore, once the
oxygen molecule has dissociated, atomic oxygen is also reactive on the
planar graphene sheet (3.23 eV compared to atomic
oxygen\cite{GLee:2009}), thus making the dynamics of the dissociated
oxygen potentially very complicated. Oxygen saturated edges were also
found to be the most stable in another study\cite{Seitsonen:2010},
where the authors also consider edge configurations beyond saturation
by a single atom.

In addition to $\mathrm{H}$ and $\mathrm{O}$, the adsorption of
$\mathrm{H_2O}$ and $\mathrm{CO}$ is also found to stabilize most
edges. In the case of $\mathrm{H_2O}$, the dissociation of the water
molecule is always energetically favoured compared to the molecular
adsorption. In particular, it dissociates into $\mathrm{OH}$ and
$\mathrm{H}$ on ac, ac(56), zz(57) and ac(677). On the zz edge,
instead, $\mathrm{H_2O}$ spontaneously dissociates into an adsorbed
oxygen at the edge and releasing the two hydrogen atoms in the gas
phase. Nitrogen species and $\mathrm{CO_2}$ bind rather weakly or not
at all on all edge configurations.

\begingroup
\squeezetable
\begin{table}
  \begin{tabular}{c|c|c|c|c|c|c|c|c|c|c}
    \hline
    \hline
    &\multicolumn{2}{|c|}{zz(57)}&\multicolumn{2}{|c|}{ac}&\multicolumn{2}{|c|}{ac(56)}&\multicolumn{2}{|c|}{zz}&\multicolumn{2}{|c}{ac(677)}\\
    &300K&0K&300K&0K&300K&0K&300K&0K&300K&0K\\
    \cline{1-11}
    \hline
    pristine& &0.92& &0.96& &1.30& &1.16& &1.04 \\
    \hline
    $\mathrm{H}$&0.46&0.30&0.21&0.03&0.75&0.66&0.25&0.10&0.61&0.43 \\
    \hline
    $\mathrm{O}$&0.00&-0.13&-0.03&-0.19&0.21&0.13&-0.22&-0.36&0.17&0.02 \\
    \hline
    $\mathrm{N}$&0.98&0.93&1.24&1.17&1.13&1.06&1.28&1.17&1.21&1.14 \\
    \hline
    $\mathrm{CO}$&0.90&0.70&0.66&0.42&0.82&0.58&0.99&0.58&1.05&0.81 \\
    \hline
    $\mathrm{N_2}$&1.22&1.11&1.20&1.07&1.14&1.00&1.47&1.24&1.12&0.99 \\
    \hline
    $\mathrm{O_2}$&0.42&0.28&0.82&0.66&0.69&0.54&1.03&0.77&1.06&0.99 \\
    \hline
    $\mathrm{H_2O}$&0.59&0.46& 0.52&0.36&0.87&0.71&1.09&0.81&0.86&0.70 \\
    \hline
    $\mathrm{CO_2}$&0.89&0.75&0.93&0.76&1.42&1.25&1.71&1.42&1.36&1.19 \\
    \hline
    \hline
  \end{tabular}
  \caption{Edge formation free-energy at $0 K$ and $300 K$ for the different nanoribbons and adsorbed species. All energies are in eV/\AA.}
  \label{tab:table2}
\end{table}
\endgroup

\begin{figure}
    \includegraphics[width=0.49\textwidth]{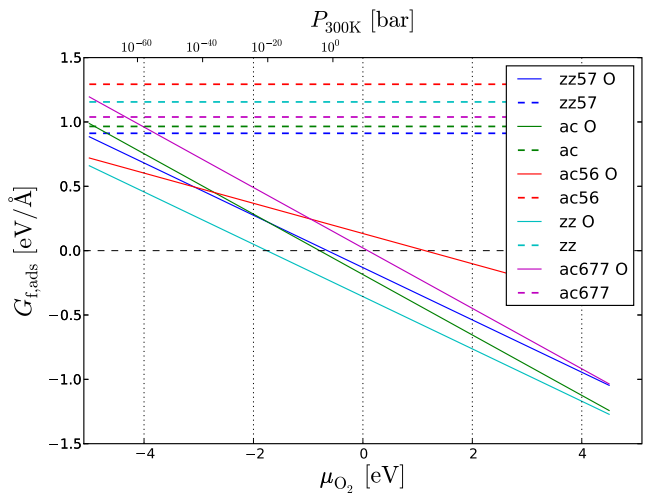}
    \caption{(Color online) Edge formation free-energy for the
      different edge configurations of a graphene nanoribbon in a
      background of $\mathrm{O_2}$ gas.}
  \label{fig:4}
\end{figure}

These calculated data can furthermore be used in order to construct
phase diagrams as a function of the chemical potential of a given gas
species. This might be relevant for guiding experimental
investigations in which GNRs are studied in the presence of a given
gas species. An example is shown in Fig. \ref{fig:4} for the case of
dissociated oxygen at the edges. This shows that for any
experimentally accessible pressure of $\mathrm{O_2}$, the most stable
edge configuration is the standard zig-zag, saturated by oxygen.

\section{Oxygen saturated zig-zag edge}
In the previous section we found that the zig-zag GNR saturated by
atomic oxygen is the most stable thermodynamically. For this reason,
we investigate in this section the ground state and vibrational
properties of the oxygen passivated zig-zag GNR and compare it to the
well studied hydrogen passivated case.

\subsection{Bandstructure}

\begin{figure}
    \includegraphics[width=0.49\textwidth]{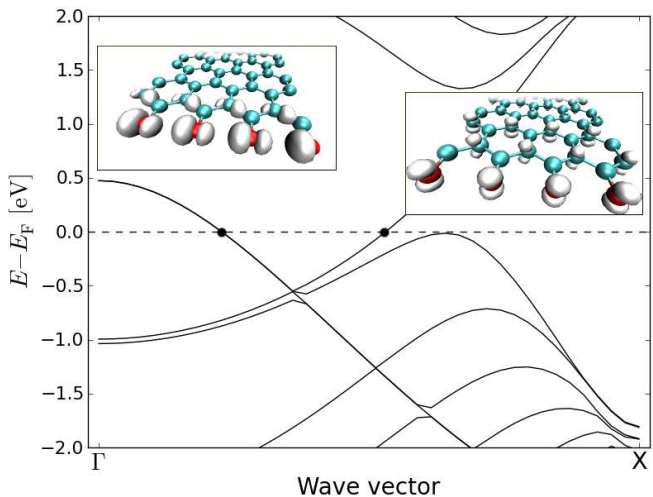}
    \caption{(Color online) Band structure of a $22\, \mathrm{\AA}$
      wide zig-zag GNR saturated with $\mathrm{O}$. The two insets show the
      states corresponding to the bands crossing the Fermi level,
      marked with black circles.}
  \label{fig:5}
\end{figure}
Fig. \ref{fig:5} shows the Kohn-Sham band structure for the oxygen
saturated GNR calculated with the RPBE functional. Contrary to the
hydrogen saturated case, its ground state is spin-paired and
metallic. The states at the Fermi level are localized at the edges,
and they have weight on the $p$ orbitals of the oxygen atom. One state
has $p_z$ symmetry and it decays into the GNR away from the edge with
weight on only one carbon sublattice (left inset of
Fig. \ref{fig:5}). The other state is almost completely localized on
the out of plane $p$ oxygen orbital (right inset of
Fig. \ref{fig:5}). This is also in constrast to the hydrogen saturated
case, where the edge states are entirely localized on the edge carbon
atoms.

\subsection{Vibrational properties}
Low energy spectroscopies probing the vibrational excitations have
traditionally been a powerful tool for structure determination in
molecules.  It has recently become possible to study differences in
the edge orientation of GNRs using Raman
spectroscopy\cite{Casiraghi:2009,Cong:2010}. In view of this we have
calculated the vibrational spectrum of the zig-zag GNRs passivated by
hydrogen and oxygen in order to identify features specific to these
types of edges.

\subsubsection{Computational details}
To obtain the phonon frequencies we first compute the dynamical matrix
in real space. The latter is obtained by numerical differentiation of
the forces.  A careful procedure to obtain the equilibrium
configuration is found essential in order to obtain an accurate
description of the phonon frequencies. Thus, for a single unit of the
zzGNR, we used a grid spacing of 0.1 \AA\ and a $k$-point sampling of
($1 \times 1 \times 10$). The relaxation of the equilibrium structure
is continued until the maximal residual forces are less than
0.02eV/\AA.

The dynamical matrix is evaluated in a supercell containing seven
primitive cells of the GNR. All atoms in one of the primitive cells
are displaced by $\pm 0.02$ \AA~in the $x,y$, and $z$ directions (a
total of $6\times N$ calculations ) and the forces on all remaining
atoms within a truncation radius of 7 \AA~of the displaced atom are
stored. This requires a total of $6\times N$ calculations where $N$ is
the number of atoms in a primitive cell.  For these calculations we
use a grid spacing of 0.18 \AA~and a $1\times 1 \times 3$ $k$-point
sampling.  We have verified that our results have converged with
respect to these parameters and the truncation radius. The real space
dynamical matrix is obtained from a finite difference of the forces,
and the eigenvectors and frequencies are obtained after Fourier
transformation.

\subsubsection{Results}
\begin{figure}[!ht]
  \includegraphics[width=0.5\textwidth]{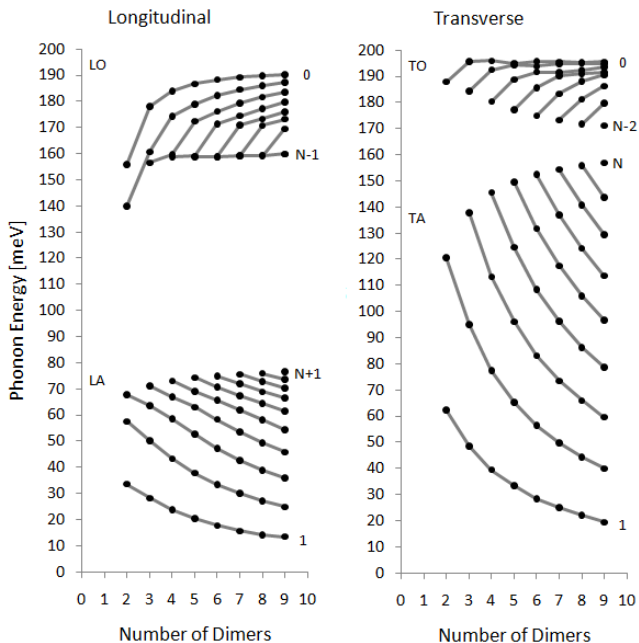}
  \caption{The longitudinal (left) and transverse (right)
    $\Gamma$-point spectrum of the oxygen saturated zigzag nanoribbon
    as function of the number of carbon dimers. The fundamental mode
    of the acoustical series is zero for all widths and is not shown.}
  \label{fig:6}
\end{figure}

The GNRs are finite in their width direction and thus they are subject
to boundary conditions similar to that of a string with free ends. The
edges therefore only allow standing waves perpendicular to the ribbon
axis and the phonon wave vector for a GNR consisting of $N$ carbon
dimers will be quantized as $q_{\perp,n} = \pi n/w$, where $w$ is the
width of the dimers and $n=0$,...,$N-1$ each corresponding to a normal
mode. The dispersion relations therefore consist of six fundamental
modes each with $(N-1)$ overtones. A study of the $\Gamma$-point
spectrum according to this classification has been done in a previous
study \cite{Gillen:2009} for hydrogen saturated nanoribbons with
purely zigzag and armchair edges. The modes can be separated into the
three categories according to their longitudinal, transverse, and
out-of-plane nature. It is straightforward to identify which category
a particular mode belongs to by weighting each of the three different
spatial entries in the polarisation vector relative to its total
norm. Each of the spatial categories can be separated into acoustical
and optical parts giving rise to six series. The $\Gamma$-point
spectrum is found to be similar to hydrogen saturated ribbons where a
splitting into optical (160-200)meV and acoustical modes (0-160)meV is
observed, as shown in Fig. \ref{fig:6}. The transverse acoustical
modes are seen to follow a near perfect $1/\sqrt{w}$ behaviour which
is understood in terms of the ribbon acting as a spring perpendicular
to its axis with fixed spring constant.

A separation of modes associated with the saturation atoms are not as
simple in the case of oxygen as it is for hydrogen. Contrary to the
hydrogen passivated ribbons where three degenerate pairs of edge
localized modes are found at 380meV, 150meV and 125meV, only one
localized degenerate pair of carbon-oxygen modes shows up, namely the
degenerated stretching modes (transverse direction). Similarly to the
hydrogen passivated case, this pair is found to be independent of the
ribbon width and is located around 182meV, except for the very narrow
ribbons ($N<4$) where it is slightly higher. The last four modes are
not separated from the remaining modes as for the case of hydrogen. In
fact, the oxygen atoms participate as a continuation of the carbon
system and enters in the bulk vibrational modes giving rise to two
additional acoustical modes in the longitudinal series and the
out-of-plane series. This coupling to the acoustical modes is
understood in terms of the comparable masses of the two species. Also,
for the transverse series one finds that an optical mode has been
converted into an additional acoustical one. The optical modes of the
longitudinal and out-of-plane series on the contrary remains similar
to what is found for the hydrogen saturated ribbons, that is, the
series are constituted of normal modes only coming from the carbon
atoms.
\begin{figure}[!ht]
  \includegraphics[width=0.5\textwidth]{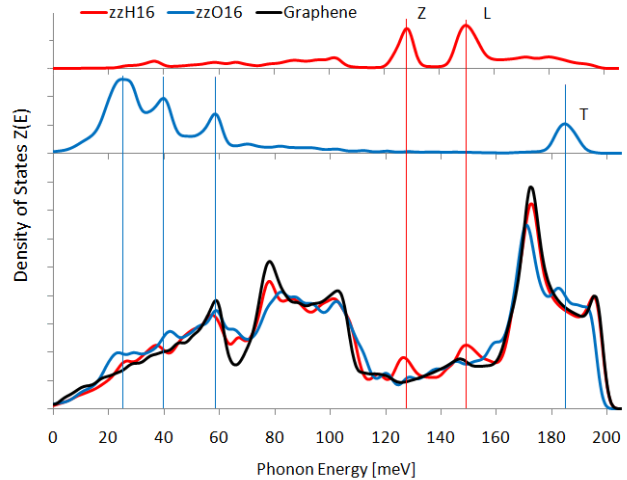}
  \caption{(Color online) The total vibrational density of state of
    the 16 dimer wide (33\AA) hydrogen(red) and oxygen(blue) saturated
    zigzag nanoribbon in comparison with graphene(black). The density
    of states weighted on the saturating atoms shown for the hydrogen
    and oxygen saturated 16 dimer zigzag ribbon, respectively.}
  \label{fig:7}
\end{figure}

The total vibrational density of states of the 16 dimer wide zigzag
ribbon is obtained by integration in the $\Gamma KM$-direction and is
shown in Fig. \ref{fig:7}. Note that it represents the whole spectrum
of states and not just the $\Gamma$-phonons. Except for some distinct
peaks the total density of states for the hydrogen and oxygen
saturated ribbons represents that of graphene fairly well. In order to
compare the deviations in the density of states associated with the
edge, the states localised on the edges have been separated out for
each of the saturating species. The peaks present in the edge
contribution are to a large degree recognisable in the corresponding
total density of states spectrum. For the hydrogen saturated ribbon,
the two peaks observed are the vibrational mode perpendicular to the
carbon-hydrogen bond and out-of-plane mode. For the oxygen saturated
ribbon only the transverse carbon-oxygen has a truly edge-localized
vibrational pattern while the peaks observed in the acoustical part of
the spectrum originates from individual normal modes in the
longitudinal and out-of-plane series.

\section{Conclusion}
As a general reminder, it should be noted that the results in this
study do not consider the effects of the substrates. In most
experiments, GNRs are grown or deposited on a substrate, which might
affect the energetics at the edges. The interaction of the substrate
with graphene edges is likely to be much stronger than the interaction
with the bulk graphene\cite{Giovannetti:2008,Vanin:2010}. 

In conclusion we have studied the adsorption of common gas molecules
at the edge of several types of GNRs using density functional theory.
We have calculated the edge formation free energy in the presence of
the adsorbates at the different edges and shown that passivation by
atomic oxygen is the most stable closely followed by the hydrogen
passivated GNRs. In contrast to the hydrogen saturated zig-zag edge,
passivation by oxygen leads to a non-magnetic and metallic
groundstate.

For the vibrational properties, the comparable masses of oxygen and
carbon lead to a larger degree of mixing between localized edge and
bulk modes. In particular, it is found that the oxygen atoms act as a
continuation of the carbon system and participates actively in the
longitudinal and out-of-plane series. Only the stretching (transverse)
mode of the carbon-oxygen bond is found to be completely localized at
the edge and separated from the normal mode series.

Since $\mathrm{O_2}$ and $\mathrm{H_2}$ are almost always present
under realistic conditions, our results suggest that it might be very
difficult to functionalize the edge atoms with different species. We
suggest that the substitutional route, for example via electrothermal
reactions as demonstrated in \cite{Wang:2009}, might turn out to be an
easier approach for doping GNRs.

\section{acknowledgments}
The authors acknowledge support from the Danish Center for Scientific
Computing. The Center for Atomic-scale
Materials Design is sponsored by the Lundbeck Foundation.

\bibliographystyle{apsrev}

\end{document}